\documentclass[graybox]{svmult}

\usepackage{helvet}         
\usepackage{courier}        
\usepackage{type1cm}        
\usepackage{makeidx}         
\usepackage{graphicx}        
\usepackage{multicol}        
\usepackage[bottom]{footmisc}

\usepackage{def} 
\definecolor{Grmy}{rgb}{0.7, 0.7 ,0.7}
\definecolor{blue}{rgb}{0, 0 ,1}

\newcommand{\blue}[1]{\color{blue}{#1}}
\newcommand{\red}[1]{\textcolor{red} {#1}}
\newcommand{\text}[1]{\mathrm{#1}}

\usepackage[utf8]{inputenc}
\usepackage{slashed}
\usepackage{amssymb} 
\usepackage{stackengine} 
\usepackage{amsmath}

\usepackage[square,comma,numbers,sort&compress]{natbib}

\begin{document}

\title*{The chirality-flow formalism for standard model calculations}
\author{Joakim Alnefjord, Andrew Lifson, Christian Reuschle and Malin Sj\"odahl}
\institute{
  {Joakim Alnefjord, Andrew Lifson, Christian Reuschle and Malin Sj\"odahl (speaker),  \at Department of Astronomy and Theoretical Physics, Box 43, 221 00 Lund, Sweden, \email{malin.sjodahl@thep.lu.se}
}}
\maketitle

\abstract{
  Scattering amplitudes are often split up into their color ($\mathfrak{su}(N)$) and kinematic
  components.
  Since the $\mathfrak{su}(N)$ gauge part can be described using flows of color,
  one may anticipate that 
  the $\mathfrak{su}(2)\oplus \mathfrak{su}(2)$ kinematic part
  can be described in terms of flows of chirality. In two recent papers we showed that this
  is indeed the case, introducing the chirality-flow formalism for standard model
  calculations.  Using the chirality-flow method --- which builds on and further
  simplifies the spinor-helicity formalism --- Feynman diagrams can be directly
  written down in terms of
  Lorentz-invariant spinor inner products, allowing the simplest and most direct
  path from a Feynman diagram to a complex number. In this presentation, we introduce
  this method and show some examples.}

\section{Introduction}
\label{sec:intro}

Since a few decades it is known that calculations in SU(3) color space 
can be elegantly simplified using a flow picture for color
\cite{tHooft:1973alw,Maltoni:2002mq}.
In this talk we ask the question if we can similarly simplify the
Lorentz structure, which at the algebra level is associated with
a left and a right chiral $\mathfrak{su}(2)$.

More specifically, bearing in mind that for color, one can formulate color-flow Feynman
rules, we ask whether we can analogously formulate a set of chirality-flow Feynman rules
to simplify calculations of Lorentz structure.
In this presentation we will answer this question affirmatively
and show how Feynman rules can be recast into chirality flows
and that this beautifully simplifies calculations \cite{Lifson:2020oll,Lifson:2020pai,Alnefjord:2020xqr,Alnefjord:2021yks}.
    
On the QCD side, we can translate every color structure to flows of
color using the $\mathfrak{su}($N$)$ Fierz identity to remove adjoint indices
\begin{eqnarray}
  \label{eq:QCDFierz}
  \underbrace{\raisebox{-0.3\height}{\includegraphics[scale=0.35]{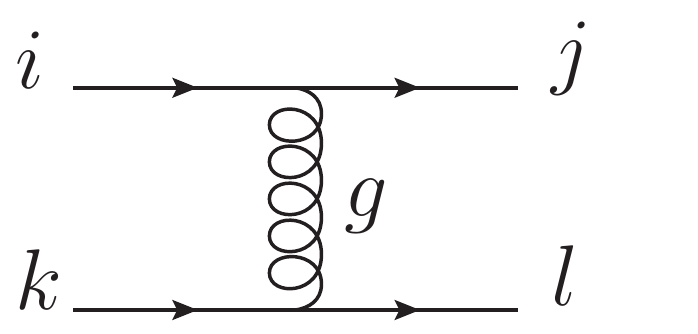}}}_{t^g_{ij}t^g_{kl}}
  &=&
  \underbrace{\raisebox{-0.3\height}{\includegraphics[scale=0.35]{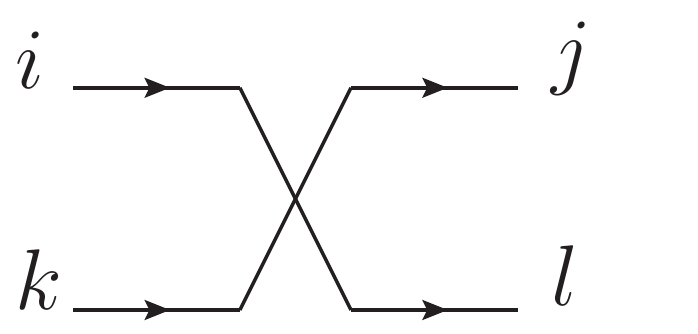}}}_{\delta_{il} \delta_{kj}}
  -\frac{1}{N}
  \underbrace{\raisebox{-0.3\height}{\includegraphics[scale=0.35]{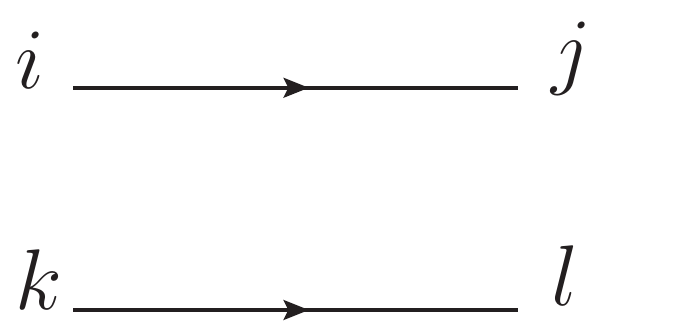}}}_{\delta_{ij}\delta_{kl}}.
\end{eqnarray}
Similarly, external gluons can be rewritten in terms of color-anticolor
pairs (with a color suppressed ``U(1)'' gluon contribution),
and the color structure of triple-gluon vertices can be expressed in
terms of traces, such that in the end, every amplitude is a linear combination of products of
Kronecker deltas in color space \cite{tHooft:1973alw,Maltoni:2002mq}.

Before attempting the same procedure for the Lorentz structure, we
recall that
at the level of the (complexified) algebra, the Lorentz
group consists of two copies of $\mathfrak{su}(2)$,
$\textcolor{blue}{\mathfrak{su}(2)_{\text{left}}} \oplus \textcolor{red}{\mathfrak{su}(2)_{\text{right}}}$,
and that the Dirac spinor structure transforms under the direct sum representation
$(\textcolor{blue}{\frac{1}{2}},0)\oplus (0,\textcolor{red}{\frac{1}{2}})$.
In the chiral (or Weyl) basis we have (for some conventions)
\begin{equation}
  \begin{pmatrix}
    {\blue{u_L}}\\
    {\red{u_R}}
  \end{pmatrix}
  \rightarrow 
  \begin{pmatrix}
    \blue{e^{-i\bar{\theta} \cdot \frac{\bar{\sigma}}{2}+\bar{\eta} \cdot \frac{\bar{\sigma}}{2}}} & 0\\
    0 &  \red{e^{-i \bar{\theta} \cdot \frac{\bar{\sigma}}{2}-\bar{\eta} \cdot \frac{\bar{\sigma}}{2}}}
  \end{pmatrix}
  \begin{pmatrix}
    {\blue{u_L}}\\
    {\red{u_R}}
  \end{pmatrix}\;,
\end{equation}
i.e. we actually have two copies of SL(2,$\mathbb{C}$), generated by the
complexified $\mathfrak{su}(2)$ algebra.

We will build heavily on the chiral representation and the
spinor-helicity formalism
\cite{DeCausmaecker:1981jtq,Berends:1981uq,Berends:1983ez,Kleiss:1984dp,Berends:1984gf,Gunion:1985bp,Gunion:1985vca,Hagiwara:1985yu,Kleiss:1986qc,Xu:1986xb}, and start with considering the
massless case, for which
\begin{eqnarray}
  u^+(p) =
  \begin{pmatrix}
    0 \\
    \ranSp{p}
  \end{pmatrix}, 
  u^-(p) =
  \begin{pmatrix}
    \sqrSp{p} \\
    0
  \end{pmatrix},
  \ubar^+(p) =
  \begin{pmatrix}
    \sqlSp{p}, & 0
  \end{pmatrix},
  \ubar^-(p) =
  \begin{pmatrix}
    0, & \lanSp{p}
  \end{pmatrix}\;.
\end{eqnarray}
From the spinor-helicity formalism
we also borrow the expressions
for the polarization vectors \cite{Xu:1986xb,Gunion:1985vca}, expressed in terms
of the physical momentum $p$, and a reference momentum $r$
\begin{equation}
  \eps_{\blue{L}}^{\mu}(p,r) \rightarrow \frac{\ranSp{r} \sqlSp{p}}{\lan r p \ran}\text{  or  } \frac{\sqrSp{p} \lanSp{r} }{\lan r p \ran}, 
  \qquad \eps_{\red{R}}^{\mu}(p,r) \rightarrow \frac{\sqrSp{r} \lanSp{p}}{\sql pr\sqr } \text{  or  } \frac{\ranSp{p} \sqlSp{r} }{\sql pr\sqr }\;,
\end{equation}
where $\eps_{\blue{L}}$ is for incoming negative helicity or outgoing positive helicity
and  $\eps_{\red{R}}$  is for incoming positive helicity or outgoing negative helicity.

To construct Lorentz invariant amplitudes we build invariant spinor inner products
using the only SL(2,$\mathbb{C}$) invariant tensor, $\epsilon^{\alpha \beta}$
($\epsilon^{12} = -\epsilon^{21} = \epsilon_{21} = -\epsilon_{12}=1$).
With $\lanSp{i}=\lanSp{p_i}$ etc., we have
\begin{equation}
  \underbrace{\epsilon^{\al \be } \ranSp{i}_{\be}}_{\equiv \lanSp{i}^{\al}}\ranSp{j}_{\al}
  =\lanSp{i}^{\al}\ranSp{j}_{\al}
  =\lan i j \ran,
  \quad 
  \underbrace{\epsilon_{\da \db} \sqrSp{i}^{\db}}_{\equiv \sqlSp{i}_{\da}} \sqrSp{j}^{\da}=
  \sqlSp{i}_{\da} \sqrSp{j}^{\da}
  =\sql ij\sqr\;.
\end{equation}
Amplitudes are thus built up out of contractions of the form
$\lan i j \ran, \sql ij \sqr\sim\sqrt{s_{ij}}$,
and if we manage to create a flow picture, the ``flow'' must contract
left (dotted) and right (undotted) indices separately.

\section{Towards chirality flow}

For the Lorentz structure, a fermion-photon vertex is associated with a factor
$\gamma^{\mu}= \sqrt{2}\begin{pmatrix} 0 & \tau^\mu\\ \taubar^\mu & 0 \end{pmatrix}$
($\tau^{\mu}=\sigma^\mu/\sqrt{2}$ normalized in analogy with eq.~\eqref{eq:QCDFierz}).
This can be split into two terms, and when a $\tau^{\mu}$ from one vertex 
is contracted with a $\taubar^{\mu}$ from another other vertex, we have (always reading indices along arrows),
\begin{equation}
     \underbrace{\raisebox{-0.42\height}{\includegraphics[scale=0.35]{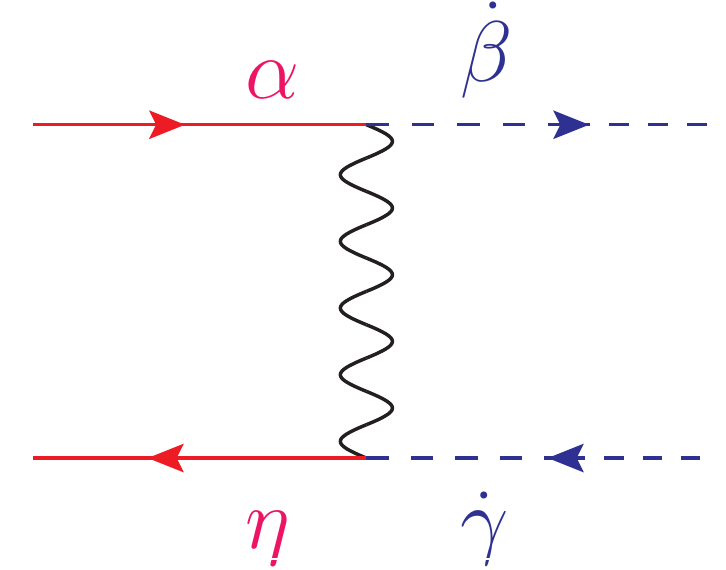}}}_{\taubar^{\mu}_{\al \db}\tau_{\mu}^{\dg \cet}}  = 
     \underbrace{\raisebox{-0.42\height}{\includegraphics[scale=0.35]{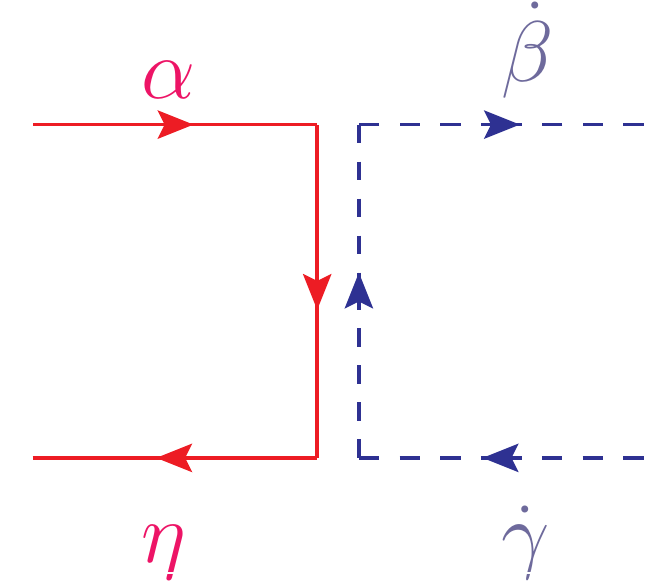}}}_{\delta_{\al}^{~\cet} \delta_{~\db}^{\dg}}\;.
\end{equation}
We note that due to the presence of $\tau^0$ there is no $1/N$-suppressed
term. In this sense chirality flow is even simpler than color flow.

When a $\tau$($\taubar$) is contracted with a $\tau$($\taubar$) from the other vertex,
the situation is more subtle, and we have to apply charge conjugation
\textit{at the level of expressions contracted with spinors} 
before removing the vector index
\begin{align*}
  \underbrace{\raisebox{-0.4\height}{\includegraphics[scale=0.4]{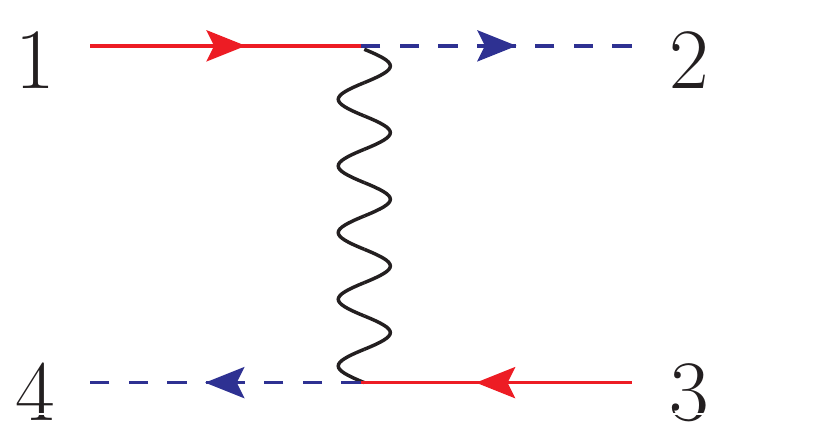}}}
  _{(\lanSp{1}^{\al}\Bar{\tau}^{\mu}_{\al \db}\sqrSp{2}^{\db}) (\lanSp{3}^{\cga}\Bar{\tau}_{\mu,\cga \deta}\sqrSp{4}^{\deta})}
  = \underbrace{\raisebox{-0.4\height}{\includegraphics[scale=0.4]{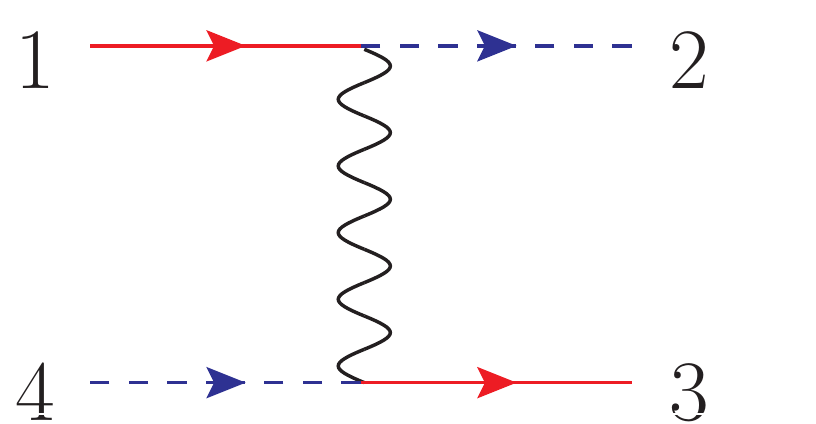}}}
  _{(\lanSp{1}^{\al}\Bar{\tau}^{\mu}_{\al \db}\sqrSp{2}^{\db}) \underbrace{({\sqlSp{4}}_{\deta}\tau_{\mu}^{\deta\cga}\ranSp{3}_{\cga})}_{\text{charge conjugated}}} = 
  \underbrace{\raisebox{-0.4\height}{ \includegraphics[scale=0.4]{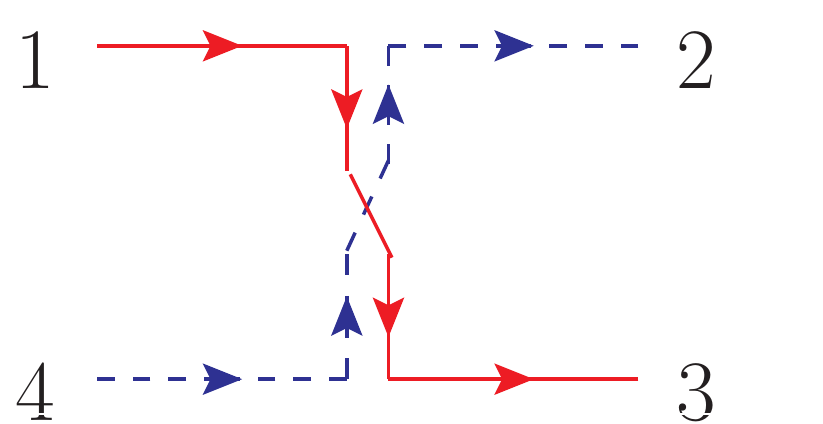}}}
  _{\lan 1 3 \ran \sql 42 \sqr}\;,
\end{align*}
were we have implicitly used the identification of spinors and their
graphical representation
\begin{eqnarray}
  \sqrSp{j}=\raisebox{-0.3\height}{\includegraphics[scale=0.3]{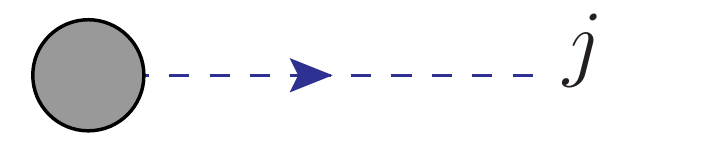}}
  \hspace*{-.3 cm},
  \sqlSp{i}= \raisebox{-0.3\height}{\includegraphics[scale=0.3]{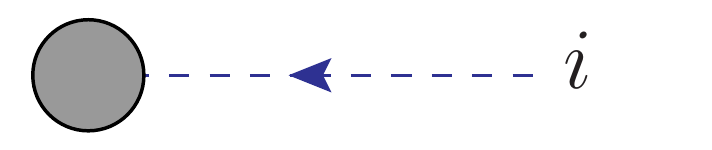}}
  \hspace*{-.3 cm},
  \ranSp{j}=\raisebox{-0.3\height}{\includegraphics[scale=0.3]{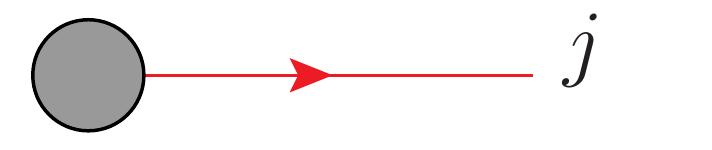}}
  \hspace*{-.3 cm},
  \lanSp{i}= \raisebox{-0.3\height}{\includegraphics[scale=0.3]{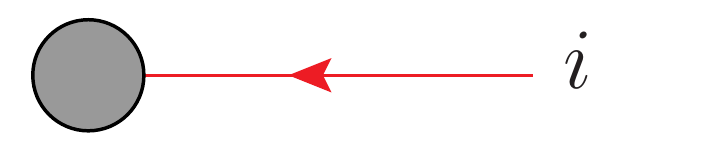}}
   \hspace*{-.3 cm}.\nonumber
\end{eqnarray}

In a similar way, charge conjugation can be applied when additional
photons are attached to a quark-line \cite{Lifson:2020pai}. A consistent arrow direction
with opposing arrows for spin-1 particles can therefore always be chosen \cite{Lifson:2020pai},
and the fermion-photon vertex can be translated to
\begin{eqnarray}
  \raisebox{-0.35\height}{\includegraphics[scale=0.35]{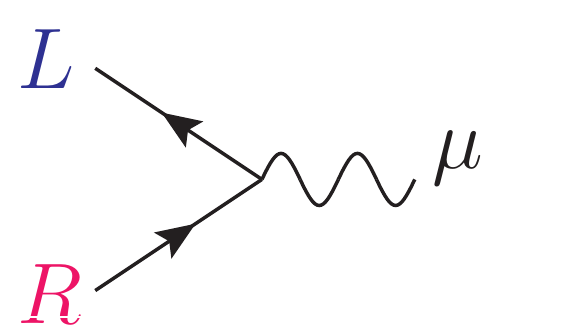}}
  \rightarrow ie\sqrt{2} \raisebox{-0.35\height}{\includegraphics[scale=0.3]{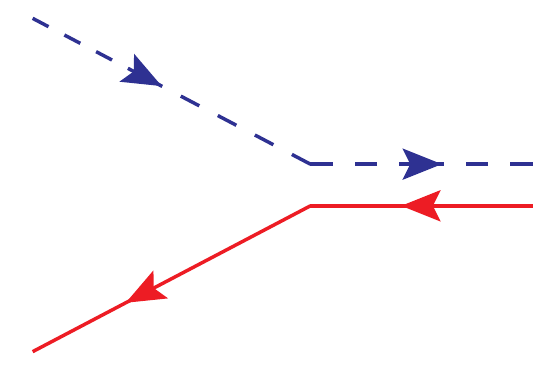}}\;, \quad
  \raisebox{-0.35\height}{\includegraphics[scale=0.35]{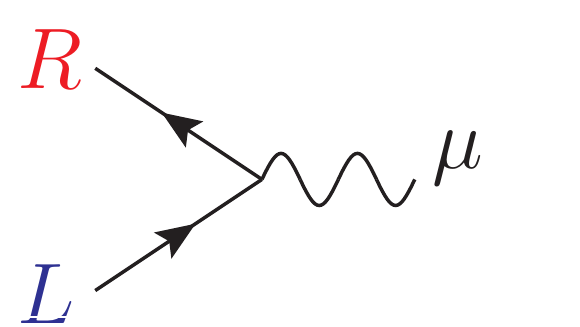}}
  \rightarrow  ie\sqrt{2} \raisebox{-0.35\height}{\includegraphics[scale=0.3]{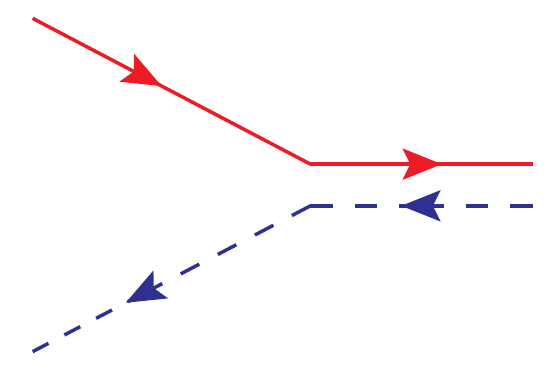}}\;.\nonumber
\end{eqnarray}

We also need to recast Fermion propagators to the flow picture. To this end, we split
$p_{\mu}\gamma^{\mu}= p_\mu \sqrt{2}\begin{pmatrix} 0 & \tau^\mu\\ \taubar^\mu & 0 \end{pmatrix}$ into two terms
\begin{eqnarray}
  \slashed{p} \equiv \sqrt{2}p^{\mu}\tau_{\mu}^{\da\be} = 
  \raisebox{-0.25\height}{\includegraphics[scale=0.4]{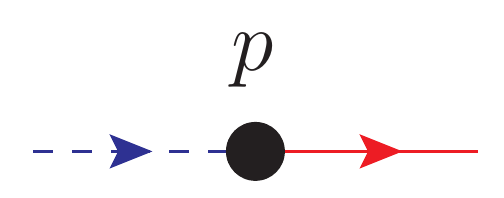}},
  \quad
  \bar{\slashed{p}}\equiv\sqrt{2}p_{\mu}\taubar^{\mu}_{\al\db} =
  \raisebox{-0.25\height}{\includegraphics[scale=0.4]{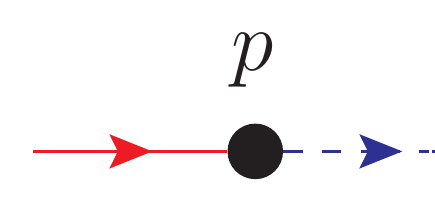}} \;,
\end{eqnarray}
where we have introduced a graphical ``momentum-dot'' notation for momenta slashed
with $\sigma$ or $\sigmabar$.

We further note that for massless momenta we have
\begin{eqnarray}
\sqrt{2}p^\mu\tau_\mu \equiv \slashed{p} = \sqrSp{p}\lanSp{p} ~, 
\quad  
\sqrt{2}p^\mu\taubar_\mu \equiv \bar{\slashed{p}} = \ranSp{p}\sqlSp{p}\;.
\end{eqnarray}
Thus any sum of light-like momenta, $p^{\mu} = \sum p^{\mu}_i \ , \ p_i^2=0$,
can be written
\begin{equation}
  \slashed{p}={\raisebox{-0.25\height}{\includegraphics[scale=0.4]{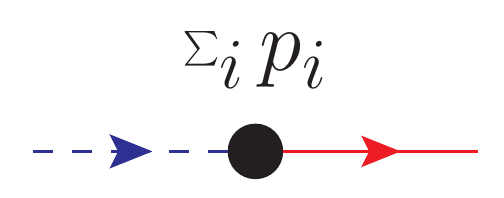}}} 
  = \sum_i \sqrSp{i}^{\da}\lanSp{i}^{\be},\; 
  \bar{\slashed{p}}= \raisebox{-0.25\height}{\includegraphics[scale=0.4]{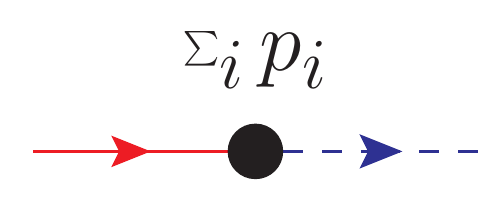}}
  = \sum_i \ranSp{i}_{\al}\sqlSp{i}_{\db}
  \;
  \text{for} 
  \;
  p_i^2=0\;.
  \nonumber 
\end{equation}
In particular, this gives for the fermion propagator
\begin{equation}
  \raisebox{-0.25\height}{\includegraphics[scale=0.4]{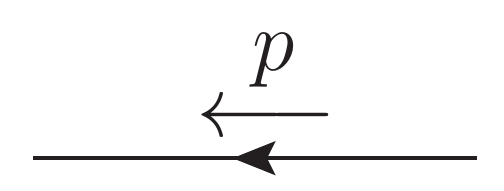}}
  \rightarrow
  \frac{i}{p^2}
  \raisebox{-0.25\height}{\includegraphics[scale=0.4]{Jaxodraw/MS/FermionPropFlowhCol.pdf}}\;\; \mathrm{or}\;\;\frac{i}{p^2}
  \raisebox{-0.25\height}{\includegraphics[scale=0.4]{Jaxodraw/MS/FermionPropFlowgCol.pdf}} \;,
\end{equation}
where the momentum is read along the fermion arrow. (It may be aligned or
anti-aligned with the chirality-flow arrows, any arrow assignment with
opposing gauge boson arrows will do for massless tree-level QED and QCD since there is always an
even number of spinor contractions \cite{Lifson:2020pai}).

For the photon propagator we have \cite{Lifson:2020pai}
\begin{equation}
  \raisebox{-0.25\height}{\includegraphics[scale=0.35]{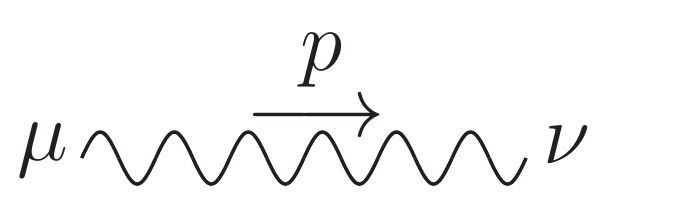}}
  \rightarrow
  - \frac{i}{p^2} \ \raisebox{-0.25\height}{\includegraphics[scale=0.4]{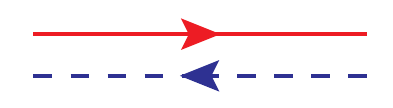}} \;
  \mathrm{or}\;\;  - \frac{i}{p^2} \ \raisebox{-0.25\height}{\includegraphics[scale=0.4]{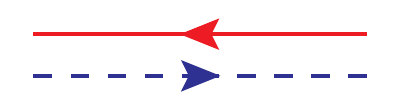}}.
\end{equation}

Finally, it is straightforward to translate the spinor structure
of external gauge bosons to the flow picture, for example 
\begin{eqnarray}
  \eps_{\blue{L}}^{\mu}(p,r) \rightarrow \frac{1}{\lan r p \ran}\raisebox{-0.2\height}{\includegraphics[scale=0.35]{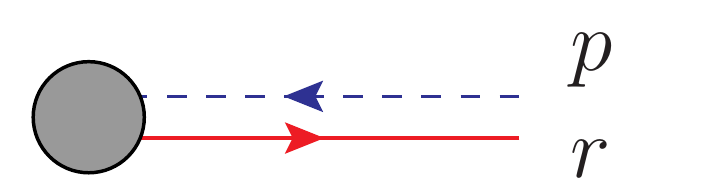}} 
  \mathrm{or}\;\;       
  \frac{1}{\lan r p \ran}\raisebox{-0.2\height}{\includegraphics[scale=0.35]{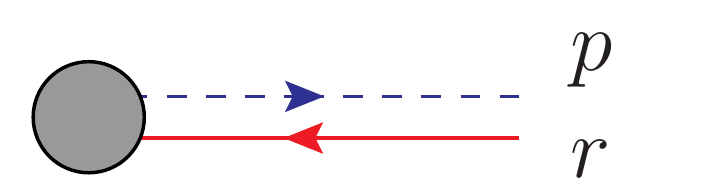}}\hspace*{-2mm}.
\end{eqnarray}

In a similar way, Feynman rules can be written down for massless QCD.
The main complication is the introduction of a momentum-dot in
the triple-gluon vertex, whereas the four-gluon vertex is just
a linear combination of chirality-flows with one dotted and one
undotted line for each metric factor \cite{Lifson:2020pai}.

\section{Examples}

Equipped with the Feynman rules for QED, we consider
the standard example of $e^+ e^-\rightarrow \mu^+ \mu^-$.
For assigned helicities, it is not hard to calculate this amplitude
within the spinor helicity formalism,
\begin{equation} 
  \raisebox{-0.4\height}{\includegraphics[scale=0.3]{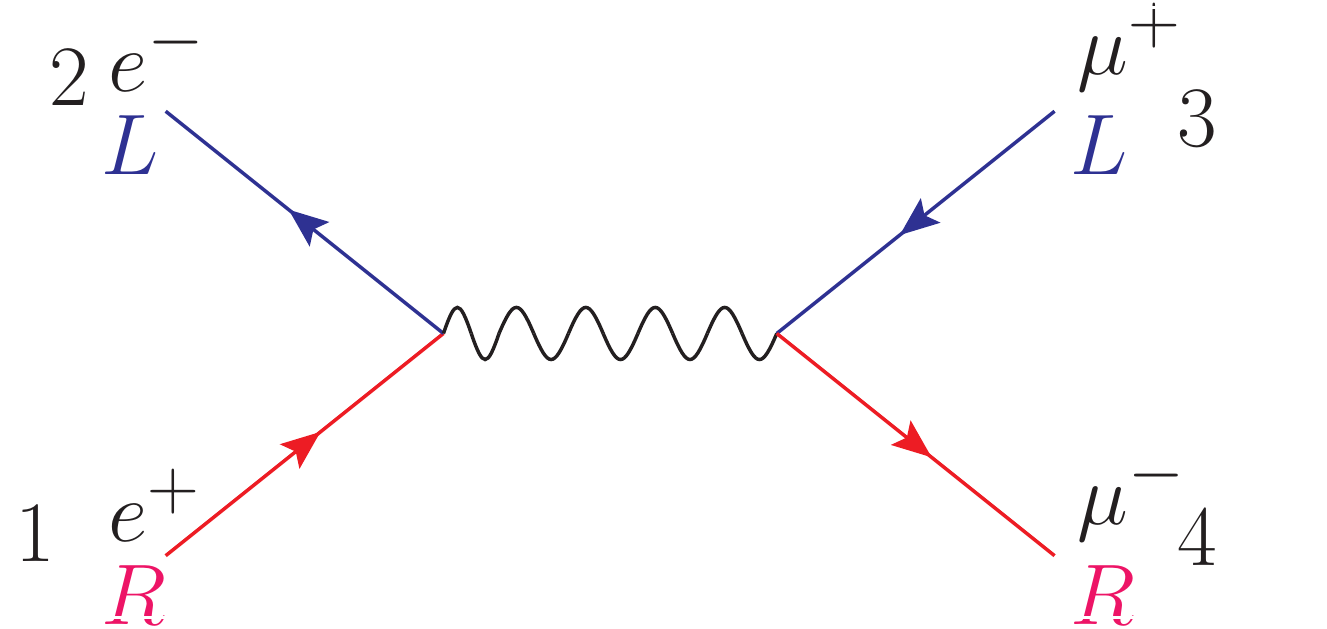}}
  \begin{tabular}{ c c }
    $=$&\hspace*{-5 mm}$\frac{2ie^2}{s_{e^+e^-}}(\sqlSp{2}_{\da}\tau_{\mu}^{\da\be}\ranSp{1}_{\be}) (\lanSp{4}^{\al}\taubar^{\mu}_{\al\db}\sqrSp{3}^{\db})\;\;\;\;\;\;\;$\\
    $=$&$\frac{2ie^2}{s_{e^+e^-}} \sqlSp{2}_{\da}\sqrSp{3}^{\da} \lanSp{4}^{\be} \ranSp{1}_{\be} = \frac{2ie^2}{s_{e^+e^-}}\sql 2\,3\sqr \lan 4\,1 \ran\;,$
  \end{tabular}
\vspace*{-0.4cm}
\end{equation}
but with chirality flow the answer can directly be drawn
\begin{eqnarray} 
  \raisebox{-0.4\height}{\includegraphics[scale=0.3]{Jaxodraw/MS/eemumuExFeynPlav3.pdf}} = 
  \frac{2ie^2}{s_{e^+e^-}}\underbrace{\raisebox{-0.4\height}{\includegraphics[scale=0.3]{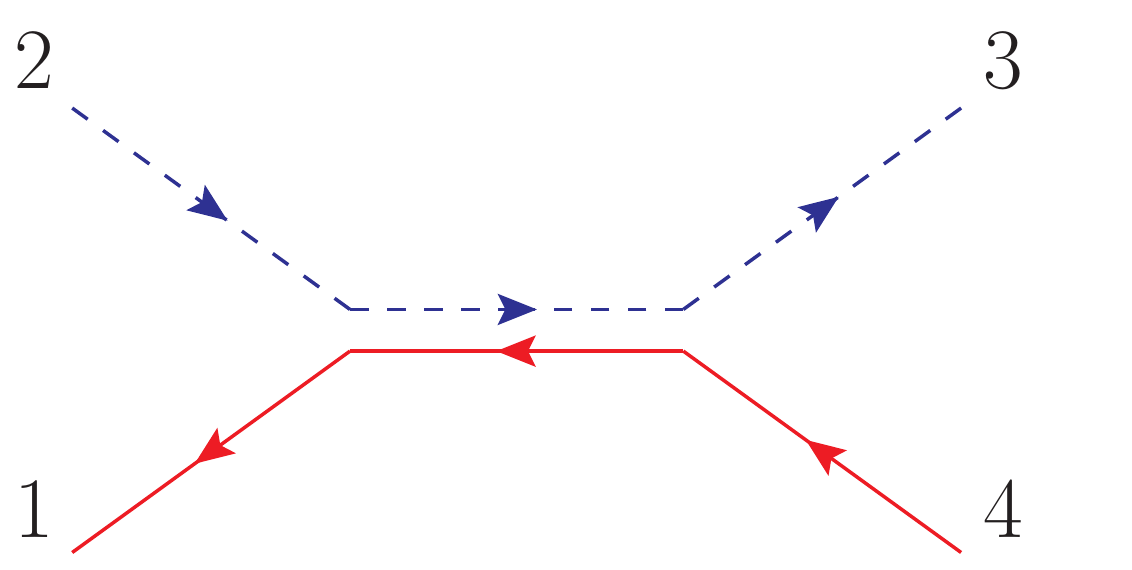}}}_{\sql 2\,3\sqr \lan 4\,1 \ran }\;.
\end{eqnarray}
\vspace*{-0.5cm}

Similarly, the value of even a very complicated massless tree-level
diagram can just be written down, for example (for reference vectors $r_8$ and $r_9$)
\begin{eqnarray}
  &&\raisebox{-0.4\height}{\includegraphics[scale=0.5]{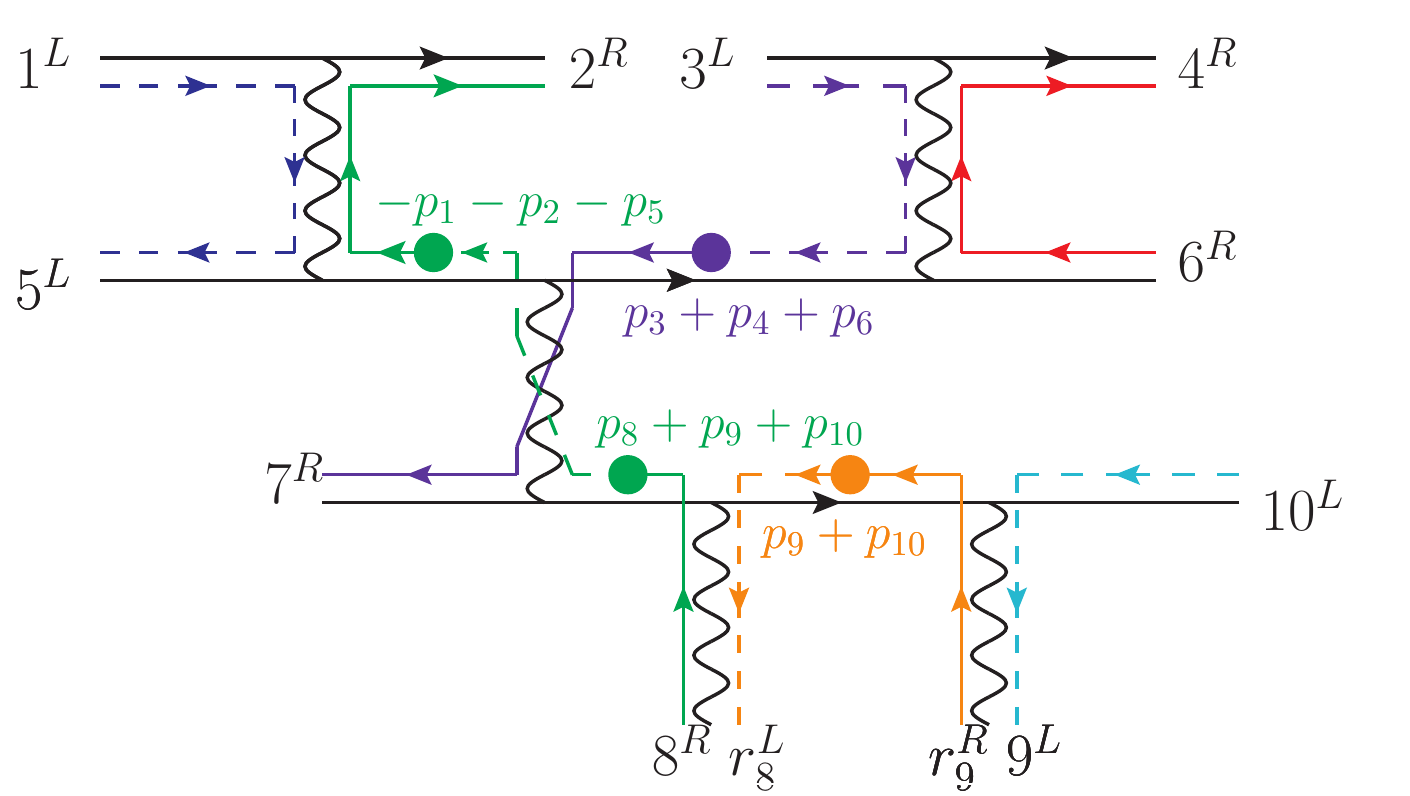}}
  \begin{tabular}{ c c}
    $=$ & $\underbrace{(\sqrt{2} e i)^8}_{\text{vertices}}\underbrace{\frac{(-i)^3}{s_{1\,2}\;s_{3\,4}\;s_{7\,8\,9\,10}}}_{\text{photon propagators}}$\\
    $\times$ & $\underbrace{\frac{(i)^4}{s_{1\,2\,5}\;s_{3\,4\,6}\;s_{8\,9\,10}\;s_{9\,10}}}_{\text{fermion propagators}}$\\
    $\times$ & $\underbrace{\frac{1}{[8 r_8 ] \langle r_9 9 \rangle}}_{\text{polarization vectors}}$
  \end{tabular}
  \nonumber\\
  &&
  \times
  {\textcolor{blue}{[ 1 5 ]}}
  {\textcolor{red}{  \langle 6 4 \rangle} }
  {\textcolor{sky}{ [ 10\,\, 9]} }   
  {\textcolor{orange} {\Bigg(
    \langle r_9 9\rangle [9r_8]
    + \langle r_9 10\rangle [10r_8]
    \Bigg)}} 
  {\textcolor{lilac}{\Bigg(
    \underbrace{[ 3 3]}_{0} \langle 3 7 \rangle
    + [3 4 ]\langle 4 7 \rangle
    + [ 3 6 ] \langle  6 7 \rangle
    \Bigg)}}
  \nonumber\\
  && \times
    {\textcolor{jaxoGreen}{\Big(
      -\langle 8 9\rangle [9 1]\langle 1 2\rangle
      -\langle 8 9\rangle [9 5]\langle 5 2\rangle
      -\langle 8 \, 10\rangle [10\,\,1]\langle 1 2\rangle
      -\langle 8 \, 10\rangle [10\,\, 5]\langle 5 2\rangle
      \Big)}}\;.
\end{eqnarray}

\section{Massive chirality flow}

To treat mass, we first note that a massive momentum $p$ always can be
written as a linear combination of two lightlike momenta, $p^{\flat}$ and $q$,
$p^{\mu} = p^{\flat,\mu} + \alpha q^{\mu}$ where $\alpha = \frac{p^2}{2p\cdot q}$.
This decomposition can be achieved in infinitely many
ways --- as is obvious from considering the system in its rest frame,
where the momenta can be taken to have any opposing direction.
Different decompositions correspond to different directions of measuring
the momentum \cite{Dittmaier:1998nn,Dreiner:2008tw,Alnefjord:2020xqr}, and in general the spin is measured along 
\begin{equation}
s^\mu = \frac{1}{m}(p^{\flat,\mu} - \alpha q^{\mu}) =
\frac{1}{m}(p^{\mu} - 2\alpha q^{\mu})\;
\end{equation}
for, for example, a $u^+$ spinor of the form \cite{Alnefjord:2020xqr}
\begin{equation}
u^{+}(p)
    =\begin{pmatrix}
    -\frac{m}{\sql q p^\flat\sqr}
    \raisebox{-0.2\height}{\includegraphics[scale=0.40]{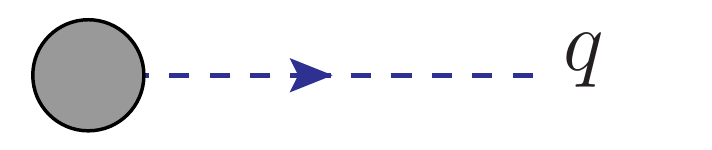}}  \\
    \phantom{-\frac{m}{\sql q p^\flat\sqr}}
    \raisebox{-0.2\height}{\includegraphics[scale=0.40]{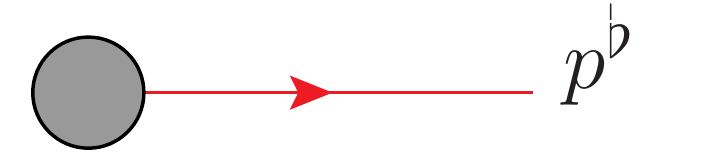}}
    \end{pmatrix}\;.
\end{equation}
The standard choice of measuring spin along the direction of motion (i.e. helicity)
corresponds to decomposing $p$ into a forward and backward direction,
$s^\mu = \frac{1}{m}(p_f^{\mu} - p_b^{\mu}) =\frac{1}{m}(|\vp|, p^0 \hat{p})$. 

Aside from massive spinors we need treat massive fermion propagators
\begin{equation}
\frac{i}{p^2-m_f^2} 
  \begin{pmatrix} 
    m_f {\delta^{\da}}_{\db} & \sqrt{2}p^{\da \be} \\ 
    \sqrt{2}\bar{p}_{\al \db} &  m_f {\delta_{\al}}^{\be}
  \end{pmatrix}
  = \frac{i}{p^2-m_f^2} 
  \begin{pmatrix} 
    m_f\includegraphics[scale=0.3]{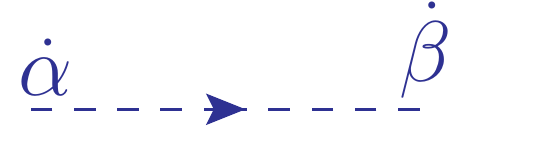}
    & \includegraphics[scale=0.3]{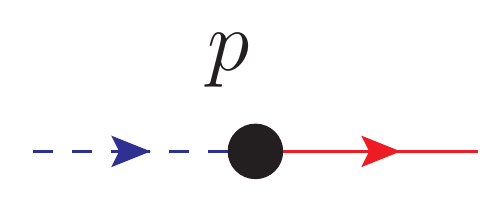} \\ 
    \includegraphics[scale=0.3]{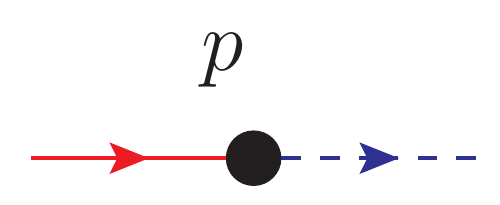}
    & m_f\includegraphics[scale=0.3]{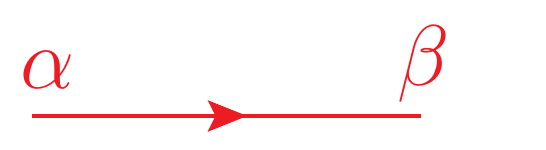}
  \end{pmatrix} \;.
\end{equation}
The presence of Kronecker delta functions may give rise to an odd
number of spinor inner products, implying that signs will have
to be carefully tracked in the massive case. We also need to
treat the third polarization degree of freedom for a massive
vector boson, but other than that, the massive case follows
quite straightforwardly from the massless case and using the
above decompositions, all Feynman rules of the standard model can be written down \cite{Alnefjord:2020xqr}.

\vspace*{-0.4cm}
\section{Conclusion}

Splitting Lorentz structure into {\blue{$\mathfrak{su}(2)_{\mathrm{left}}$}} and
{$\red{\mathfrak{su}(2)_{\mathrm{right}}}$}, we have been
able to recast all standard model Feynman rules to chirality-flow rules,
giving a transparent and intuitive way of understanding the Lorentz inner
products appearing in amplitudes.

If the ordinary spinor helicity method takes us from $4\times4$ Dirac matrices
to  $2\times2$ Pauli matrices, the chirality-flow method takes us from
Pauli matrices to scalars.
This significantly simplifies calculations with Feynman diagrams.
Many processes are within range of quick pen and paper calculations,
often without intermediate steps and the final result is transparent and intuitive.

More practically, we expect our method to be useful for event simulations with Monte
Carlo event generators, in particular when sampling over helicity.
Work towards consistent loop calculations is ongoing.

\bibliographystyle{ieeetr}
\bibliography{chiralityflow_lietheory}

\begin{thebibliography}{10}

\bibitem{tHooft:1973alw}
G.~'t~Hooft, ``{A Planar Diagram Theory for Strong Interactions},'' {\em Nucl.
  Phys.}, vol.~B72, p.~461, 1974.
\newblock [,337(1973)].

\bibitem{Maltoni:2002mq}
F.~Maltoni, K.~Paul, T.~Stelzer, and S.~Willenbrock, ``{Color Flow
  Decomposition of QCD Amplitudes},'' {\em Phys. Rev.}, vol.~D67, p.~014026,
  2003.

\bibitem{Lifson:2020oll}
A.~Lifson, C.~Reuschle, and M.~Sj\"odahl, ``{Introducing the Chirality-flow
  Formalism},'' {\em Acta Phys. Polon. B}, vol.~51, no.~6, pp.~1547--1557,
  2020.

\bibitem{Lifson:2020pai}
A.~Lifson, C.~Reuschle, and M.~Sjodahl, ``{The chirality-flow formalism},''
  {\em Eur. Phys. J. C}, vol.~80, no.~11, p.~1006, 2020.

\bibitem{Alnefjord:2020xqr}
J.~Alnefjord, A.~Lifson, C.~Reuschle, and M.~Sjodahl, ``{The chirality-flow
  formalism for the standard model},'' {\em Eur. Phys. J. C}, vol.~81, no.~4,
  p.~371, 2021.

\bibitem{Alnefjord:2021yks}
J.~Alnefjord, A.~Lifson, C.~Reuschle, and M.~Sjodahl, ``{A Brief Look at the
  Chirality-Flow Formalism for Standard Model Amplitudes},'' {\em PoS},
  vol.~LHCP2021, p.~160, 2021.

\bibitem{DeCausmaecker:1981jtq}
P.~De~Causmaecker, R.~Gastmans, W.~Troost, and T.~T. Wu, ``{Multiple
  Bremsstrahlung in Gauge Theories at High-Energies. 1. General Formalism for
  Quantum Electrodynamics},'' {\em Nucl. Phys.}, vol.~B206, pp.~53--60, 1982.

\bibitem{Berends:1981uq}
F.~A. Berends, R.~Kleiss, P.~De~Causmaecker, R.~Gastmans, W.~Troost, and T.~T.
  Wu, ``{Multiple Bremsstrahlung in Gauge Theories at High-Energies. 2. Single
  Bremsstrahlung},'' {\em Nucl. Phys.}, vol.~B206, pp.~61--89, 1982.

\bibitem{Berends:1983ez}
F.~A. Berends, R.~Kleiss, P.~de~Causmaecker, R.~Gastmans, W.~Troost, and T.~T.
  Wu, ``{Multiple Bremsstrahlung in Gauge Theories at High-energies. 3. Finite
  Mass Effects in Collinear Photon Bremsstrahlung},'' {\em Nucl. Phys.},
  vol.~B239, pp.~382--394, 1984.

\bibitem{Kleiss:1984dp}
R.~Kleiss, ``{The Cross-section for $e^+ e^- \to e^+ e^- e^+ e^-$},'' {\em
  Nucl. Phys.}, vol.~B241, p.~61, 1984.

\bibitem{Berends:1984gf}
F.~A. Berends, P.~H. Daverveldt, and R.~Kleiss, ``{Complete Lowest Order
  Calculations for Four Lepton Final States in electron-Positron Collisions},''
  {\em Nucl. Phys.}, vol.~B253, pp.~441--463, 1985.

\bibitem{Gunion:1985bp}
J.~F. Gunion and Z.~Kunszt, ``{Four jet processes: gluon-gluon scattering to
  nonidentical quark - anti-quark pairs},'' {\em Phys. Lett.}, vol.~159B,
  p.~167, 1985.

\bibitem{Gunion:1985vca}
J.~F. Gunion and Z.~Kunszt, ``{Improved Analytic Techniques for Tree Graph
  Calculations and the G g q anti-q Lepton anti-Lepton Subprocess},'' {\em
  Phys. Lett.}, vol.~161B, p.~333, 1985.

\bibitem{Hagiwara:1985yu}
K.~Hagiwara and D.~Zeppenfeld, ``{Helicity Amplitudes for Heavy Lepton
  Production in e+ e- Annihilation},'' {\em Nucl. Phys.}, vol.~B274, pp.~1--32,
  1986.

\bibitem{Kleiss:1986qc}
R.~Kleiss and W.~J. Stirling, ``{Cross-sections for the Production of an
  Arbitrary Number of Photons in Electron - Positron Annihilation},'' {\em
  Phys. Lett.}, vol.~B179, pp.~159--163, 1986.

\bibitem{Xu:1986xb}
Z.~Xu, D.-H. Zhang, and L.~Chang, ``{Helicity Amplitudes for Multiple
  Bremsstrahlung in Massless Nonabelian Gauge Theories},'' {\em Nucl. Phys.},
  vol.~B291, pp.~392--428, 1987.

\bibitem{Dittmaier:1998nn}
S.~Dittmaier, ``{Weyl-van der Waerden formalism for helicity amplitudes of
  massive particles},'' {\em Phys. Rev.}, vol.~D59, p.~016007, 1998.

\bibitem{Dreiner:2008tw}
H.~K. Dreiner, H.~E. Haber, and S.~P. Martin, ``{Two-component spinor
  techniques and Feynman rules for quantum field theory and supersymmetry},''
  {\em Phys. Rept.}, vol.~494, pp.~1--196, 2010.

\end{thebibliography}

\end{document}